\documentclass[twocolumn]{article}
\usepackage{graphicx}
\usepackage[htt]{hyphenat}

\makeatletter
\def\@normalsize{\@setsize\normalsize{12pt}\xpt\@xpt
\abovedisplayskip 10pt plus2pt minus5pt\belowdisplayskip \abovedisplayskip
\abovedisplayshortskip \z@ plus3pt\belowdisplayshortskip 6pt plus3pt
minus3pt\let\@listi\@listI} 

\def\subsize{\@setsize\subsize{12pt}\xipt\@xipt}

\def\section{\@startsection {section}{1}{\z@}{24pt plus 2pt minus 2pt}
{12pt plus 2pt minus 2pt}{\large\bf}}

\def\subsection{\@startsection {subsection}{2}{\z@}{12pt plus 2pt minus 2pt}
{12pt plus 2pt minus 2pt}{\subsize\bf}}
\makeatother

\begin{document}



\title{Making refactoring decisions in large-scale Java systems: an empirical stance} 

\author{
	Richard Wheeldon and Steve Counsell \\
	School of Computer Science and Information Systems\\
	Birkbeck University of London\\
	Malet Street,
	London,
	WC1E 7HX. \\
	Email:\{richard,steve\}@dcs.bbk.ac.uk
}

\maketitle

\subsection*{\centering Abstract}
{\em
Decisions on which classes to refactor are fraught with difficulty.
The problem of identifying candidate classes becomes acute when
confronted with large systems comprising hundreds or thousands of classes.
In this paper, we describe a metric by which {\em key} classes, 
and hence candidates for refactoring, can be identified.
Measures quantifying the usage of two forms of coupling, inheritance and
aggregation, together with two other class features (number of methods and
attributes) were extracted from the source code of three large Java systems. Our
research shows that metrics from other research domains can be adapted to
the software engineering process. Substantial differences were found
between each of the systems in terms of the key classes identified
and hence opportunities for refactoring those classes varied between
those systems.
} 

\section{Introduction}

The term {\em refactoring} refers to a technique for improving code
quality by making changes to the internal structure 
of the software without changing its external behaviour.
Refactoring can be used to improve software design by, for example,
moving code between classes, extracting code into new methods or classes
or altering the position of classes in an inheritance hierarchy.
Refactoring therefore leads the programmer to work more deeply on
understanding what the code does and is thus an aid to maintenance 
and reuse \cite{JOHN88}.

The potential benefits of carrying out refactoring are reduced
duplication of code, improved readability, faster development and
fewer bugs. Whilst there has been considerable interest 
in refactoring principles, relatively little research seems to have
focused on the identification of candidate classes for the refactoring
process itself. 

Extreme programming (XP) practitioners use refactoring to remove duplication
whenever program features are added. Beck's advice is that programmers
should not ``refactor on speculation \ldots refactor when the system asks
you to'' \cite{BECK00}. Whilst this advice may be appropriate for XP projects,
refactoring of {\em key} classes may also be required when developing libraries
for other programmers to use or when developing software in large teams.

We conjecture that there are certain classes in every OO system
which are of such importance (in terms of the features they possess) that they
should be a priority for refactoring effort. 
In this paper we show how these classes may be identified through two countable
measures (number of methods and 
number of attributes) and through a Web-based graph metric which identifies 
the extent of each class' coupling via inheritance and aggregation.
The {\em potential gain} (PG) metric was first used as a means of identifying
key web pages in a Web search and navigation system. The potential gain has been
used here to rank classes in terms of their coupling patterns.
By ranking classes according to all these features, candidate key classes
can be identified.

\section{Motivation and related work}

The motivation for the research in this paper stems from a number of sources. 
Firstly, developers will inevitably want to quickly identify which classes  
are key to the system as a whole. This may be because they wish to 
avoid applying refactoring effort to such classes, and the associated 
problems of re-testing. Alternatively, it may be that those classes are the
ones which need to be maintained most rigorously because they {\em are} 
key classes and exert considerable influence in the system as a whole. The
research in this paper allows those classes to be identified more easily.

Another motivation for the work in this paper stems from a lack of research
into refactoring decision criteria. The basis on which a particular class (or
refactoring) should be chosen is not well understood or documented. We view it
as a fruitful and interesting research topic. While there has been plenty of
evidence  for the benefits of refactoring, exactly what to refactor (and when)
is still an open issue. Availability of tool support for refactoring (although 
improving significantly) seems dwarfed by current progress and thinking in the
area. Our research is an attempt to highlight the criteria by which refactoring
effort can be initially chanelled.

A further motivation for the research is to identify whether Java classes exhibit 
similar properties to their C++ counterparts. This builds on earlier work
\cite{COUN00}, where high-level class metrics were collected from the Unified
Modelling Language (UML) \cite{RUMB98} documentation of five large C++ systems,
two of which were libraries.
Two conjectures were investigated to determine features of key classes in a 
system and to investigate any differences between library-based systems and
other systems in terms of coupling. Key classes in the three
application-based systems tended to contain significant amounts of aggregation,
large numbers of methods, attributes and associations, but little inheritance.
No consistent, identifiable key features could be found in the two library-based 
systems; both showed a distinct lack of any form of coupling (including
inheritance) other than through the C++ friend facility. 

In terms of seminal refactoring literature, the PhD. work of Opdyke \cite{OPDY92} 
describes a number of software refactorings. This thesis spawned a large amount of
research in the subject. Opdyke and Johnson \cite{JOHN93} describe a study in
which they illustrate how to create abstract superclasses from other classes by 
refactoring. They decompose the operation into a set of refactoring steps, and
provide examples. They also discuss a technique that can automate these steps 
making the process of refactoring much easier.
In Johnson and Opdyke \cite{JOHN93b}, 
some common refactorings based on aggregation are reported, including 
how to convert from inheritance to aggregation and how to reorganize an
aggregate/component hierarchy. They also describe how to refine aggregations by
moving variables and functions between aggregate and component classes, and how to
move variables and functions within inheritance hierarchies. A text by
Fowler \cite{FOWL99} describes seventy-two types of refactoring and illustrates
each type with examples and UML notation.

Recent empirical work in the refactoring area and its automation is found in
\cite{TOKU01}, where 14~000 lines of code were transformed automatically where
they would otherwise have had to be coded by hand. In Najjar et al. \cite{NAJJ03},
the opportunities, benefits and problems of refactoring class constructors across 
a sample of classes from five Java systems were investigated. The refactoring used,
{\em replacing multiple constructors with creation methods} was applied to each of
a set of classes containing three or more constructors. 

\section{Empirical Evaluation}  

We address the empirical evaluation through a single conjecture, related to key
classes as follows:  

\begin{itemize}
\item[{\bf C1:}] In the type of systems investigated, a certain number of classes 
have higher numbers of methods, attributes, aggregation relationships and 
subclasses than other classes in the same type of system. These classes will,
typically, be found towards the root of an inheritance hierarchy so that 
subclasses may take advantage of the large functionality and features they offer.
\end{itemize} 

This is what we believe characterises a key class. We accept that there are
many other forms of coupling which are equally worthy of investigation, and hence
we do not claim our criteria to be definitive. In Section 5, we justify our
choice of criteria by reference to several core refactorings.
 
\subsection{Potential Gain}

The potential gain metric was originally developed to aid the selection of pages
allowing for future navigation of Web sites. By considering the number and
length of all possible paths from a given node, an estimate of the utility
or connectivity of a node could be established.
If the density of the neighbourhood of some node in a graph is high, then many
nodes are connected via short paths and the PG of that node is also higher.
In complex systems, the lengths of these paths reflect the likely
influence on connected nodes or classes.

As part of this research, a system called AutoCode was developed for indexing
Java source code. AutoCode works by using a custom {\em doclet} which extends the
Javadoc program and allows easy access to the code structure. We used the AutoCode
system to generate graphs for each of five coupling types - Inheritance, Aggregation,
Interface, Parameter Type and Return Type. Previous work described features of
these graphs \cite{WHEE03c}. This paper describes new work investigating the utility
of the potential gain metric for analysing key classes using the Inheritance
and Aggregation graphs.

The \emph{Potential Gain}, PG$(c)$, of a class, $c$, is defined as the sum for
all lengths (or depths) of the product of the fraction of all possible coupling
paths which are of length $d$ and a discounting function $f(d)$. This gives a
measure of the number of objects with which an instance of the class might
potentially interact. For example, in the case of inheritance, the potential
gain metric reflects both the breadth and depth of a hierarchy. A formal
definition of potential gain is given in appendix A.

Related to class coupling, the PG metric gives an indication of the
inter-connectivity of a class with other classes. The type of connectivity can
be defined in the software which extracts (in the case of this research) the
aggregation and inheritance relationships. For example, a high PG value for a
class at a point in the inheritance hierarchy means that the class has a
relatively large number of descendants. A very low value indicates that it 
is a leaf node - in other words, it has no subclasses.

Although the mathematics behind the potential gain is non-trivial, computation
of the values is efficient. For example, once an appropriate graph structure
had been obtained, PG values were computed for all 6~000 classes in the JDK
in less than five seconds on a desktop-class PC.

\subsection{Data collected} 

Data was collected from three large Java systems. Those systems were chosen
since they were the subject of previous research. The three systems chosen
are used extensively by developers in industrial settings.
In keeping with previous work on C++ systems, we investigate two applications
and one library-based system. As well as identifying key classes, previous work
also found distinct differences between these two types of system, together with 
a lack of use of inheritance in all types of system. While this latter feature is 
not likely to be the case in a Java setting (because of the emphasis on
inheritance in Java), our belief is that Java classes will exhibit some
contradicting features across the systems. The three systems investigated were:

\begin{enumerate}
\item {\bf JDK} - The core Java class libraries shipped with the Java Developers Kit 
(JDK) which provide implementations of common functions required for many 
programs. This contains 1~400~000 lines of code spread over 6~000 classes.

\item {\bf Tomcat} is a servlet container used in the official reference
implementation for Java Servlets and JavaServer Pages. The source code for
Jakarta Tomcat contains 150~000 lines spread over 370 classes.

\item {\bf Ant} is a Java-based build tool and behaves in a similar way
to \texttt{make} but uses XML-based configuration files defining various tasks
to be executed. The source code for Apache Ant contains 145~000 lines of code
spread over 500 classes.
\end{enumerate} 
 
\subsubsection{Metrics collected} 

The following data was collected automatically, using 
the same software which produced values for the PG metric.

\begin{enumerate} 
\item The Number of {\bf Methods} in a class, including public, protected and
private member functions.

\item The Number of Class {\bf Attributes}, including public, protected and
private declarations.

\item The {\bf Depth} or level of a class in an inheritance hierarchy where
a zero value represents the root. The metric is based on the Depth
of Inheritance Tree metric of Chidamber and Kemerer \cite{CHID91}.
\end{enumerate} 

\section{Data Analysis}

\subsection{Summary Data}

Table~\ref{tbl:summary} gives the summary data for the three systems, in terms of maximum
and median values for the three metrics collected. The JDK shows the largest
maximum values for all three metrics. The class with 254 methods is
{\tt java.awt.Component} with 80 attributes and is found at level 1 in the JDK
inheritance hierarchy. A {\tt Component} is any object having a graphical
representation and that can interact with the user.
The JDK class with 329 attributes is class {\tt xalan.templates.Constants}
with zero methods and again, was found at depth 1. Two classes
{\tt PIORB} and {\tt NSORB} reside at the lowest depth of the JDK
library (depth 8) with 41 and 2 methods, respectively.

The Tomcat and Ant systems are comparable in terms of number of methods and
attributes. The Tomcat class with 168 methods is {\tt StandardContext},
with 49 attributes and is found at depth 2. The class with 57 attributes 
was {\tt JspC} with 45 methods, found at depth 1. This class provides the shell
for the jspc compiler.

The Ant class with 89 methods is {\tt Project}, found at depth 1 with 33
attributes. The class, {\tt CBZip2InputStream} with 47 attributes has 31
methods (see Table~\ref{tbl:summary}) and is a possible candidate for refactoring.
Visual inspection of both classes revealed two bad
smells - {\em Primitive Obsession} and {\em Long Methods} \cite{FOWL99}. In
our analysis we have viewed basic system types, such as {\tt String} as
primitive.

\begin{table}[htb]
\begin{footnotesize}
\begin{center}
\begin{tabular} {|l|l|l|l|} \hline
       &Metric &Max &Median \\ \hline
JDK    &Methods  &254 &3 \\
       &Attributes  &329 &3 \\
       &Depth  &8 &2 \\
Tomcat &Methods  &168 &5 \\
       &Attributes  &57 &2 \\ 
       &Depth  &4 &1 \\
Ant    &Methods  &89 &4 \\
       &Attributes  &47 &2 \\
       &Depth  &6 &2 \\ \hline
\end{tabular}
\end{center}
\caption{\label{tbl:summary} 
Summary data for all three systems
}
\end{footnotesize} 
\end{table}

\subsection{Investigating Conjecture C1} 

Identification of key classes has significant implications for refactoring 
and ultimately, maintainability. If such classes contain significant functionality 
and are coupled to numerous other classes, then modification of those classes needs
to be made with particular care. This is particularly important if they are found
at shallow levels in an inheritance hierarchy (i.e., a low depth) since all 
subclasses need to be considered if any change is made to that class.
 
From the summary data and the previous discussion, it appears that in each of the
three systems there are classes with large numbers of methods and attributes.  
Conjecture C1 attempts to clarify the extent to which combinations of all features,
including aggregation, are found in the same classes. By aggregation relationships,
we mean classes which either {\em use} a large number of other classes or are
{\em used} by a similarly large number of other classes (or both). Hereafter, we will
call these two types of aggregation {\em normal} aggregation and {\em reverse}
aggregation, respectively.

\subsubsection{The JDK library}

Table~\ref{tbl:jdk-revagg} shows, for the JDK system, the numbers of methods
and attributes and the depths of inheritance for the top fifteen classes when
ranked in descending order according to their 
reverse aggregation PG value. A high reverse aggregation PG value implies that 
a lot of classes use the class in question. 
Classes such as {\tt PageAttributes.MediaType}, {\tt Color}
and {\tt Character.UnicodeBlock} are used in constant (static final) declarations
and are often self-referencing.

Two of the classes identified in Table~\ref{tbl:jdk-revagg} have zero attributes
and two have zero methods. The maximum depth in the inheritance hierarchy was 3 for two
classes - {\tt Vector} and {\tt javax.print.attribute.standard.MediaSizeName}.
The presence of {\tt Hashtable} and {\tt Vector} as commonly
used objects suggests that the Collections framework introduced in JDK 1.2 has
not been fully adopted within the JDK.

\begin{table*}[htb]
\begin{footnotesize}
\begin{center}
\begin{tabular} {|l|r|r|r|r|} \hline
Classname  &  Methods  &  Attributes  & Constructors & Depth \\ \hline
PageAttributes.MediaType & 0 & 223 & 1 & 2 \\
String & 58 & 7 & 12 & 1 \\
Character.UnicodeBlock & 1 & 87 & 1 & 2 \\
HTML.Attribute & 1 & 83 & 1 & 1 \\
HTML.Tag & 4 & 82 & 3 & 1 \\
MediaSizeName & 2 & 75 & 1 & 3 \\
Color & 29 & 35 & 7 & 1 \\
Object & 12 & 0 & 1 & 0 \\
CSS.Attribute & 3 & 62 & 1 & 1 \\
AccessibleRole & 0 & 56 & 1 & 2 \\
Hashtable & 24 & 14 & 4 & 2 \\
Vector & 43 & 4 & 4 & 3 \\
Class & 69 & 17 & 1 & 1 \\
TypeCode & 19 & 0 & 1 & 1 \\
ObjectStreamField & 12 & 6 & 4 & 1 \\ \hline
\end{tabular}
\end{center}
\caption{\label{tbl:jdk-revagg} 
The fifteen classes with the highest reverse aggregation PG values: JDK
} 
\end{footnotesize}         
\end{table*} 

The top fifteen classes were then ranked in descending order according
to their normal aggregation PG value.
These fifteen classes were then compared with the reverse 
aggregation classes in Table~\ref{tbl:jdk-revagg}. 
Eight classes were found to coincide with the previous fifteen found.
Table~\ref{tbl:jdk-overlap} names these classes, together with the position
they occupied in Table~\ref{tbl:jdk-revagg}.

The classes in Table~\ref{tbl:jdk-overlap} are mostly self-referencing classes
with many static references. For example, {\tt PageAttributes.MediaType}
contains a self-reference for each paper format (e.g. A4, US Letter, etc.).
Such classes induce high PG values. A similar effect was observed with
Web-search metrics. Lempel and Moran showed that metrics such as
HITS \cite{KLEI98} and PageRank \cite{PAGE98} can be influenced by a
phenomenon called the Tightly Knit Community (TKC) Effect \cite{LEMP00}.
The TKC effect has a negative influence on search results when small
groups of pages are heavily interlinked. In the analysis of program
code, the TKC effect is useful for identifying classes or networks of
classes which are tightly bound and where functionality is impossible
to extract. This is made possible by considering more than one metric
in the analysis.

A high normal aggregation PG value would imply that an instance of this
class will make many references to objects which, in turn, make many
references to other objects.
Modifying or refactoring class with high normal aggregation PG values
require careful thought and preparation, since many other classes 
may be affected by such a change.

\begin{table}[htb]
\begin{footnotesize}
\begin{center}
\begin{tabular} {|l|r|r|} \hline
Classname                & Reverse PG & PG \\ \hline
PageAttributes.MediaType & 1 & 1 \\
Character.UnicodeBlock   & 3 & 2 \\
HTML.Attribute           & 4 & 3 \\
HTML.Tag                 & 5 & 4 \\
MediaSizeName            & 6 & 5 \\
Color                    & 7 & 14 \\
CSS.Attribute            & 9 & 6 \\
AccessibleRole           & 10 & 7 \\ \hline
\end{tabular}
\end{center}
\caption{\label{tbl:jdk-overlap} 
	Eight class appear in the top fifteen classes for both
	normal and reverse aggregation PG values: JDK
}
\end{footnotesize}         
\end{table}

By considering those classes which score highly on both metrics, we can
eliminate those classes which are self-referencing. The results of
eliminating such classes leave {\tt String}, {\tt Object}, {\tt Hashtable},
{\tt Vector} and {\tt Class} clearly identifiable as extensively used, key
classes.

Table~\ref{tbl:jdk-inheritance} shows the top fifteen classes when ranked on
descending inheritance PG values.
It is interesting to note that only one class from Table~\ref{tbl:jdk-overlap}
appears in the top fifteen JDK classes from Table~\ref{tbl:jdk-inheritance}.
This class was {\tt java.lang.Object}, as might be expected; a high inheritance
PG value implies that a class has many subclasses.

\begin{table*}[htb]
\begin{footnotesize}
\begin{center}
\begin{tabular} {|l|r|r|r|r|} \hline
Classname         &  Methods  &  Attributes  & Constructors & Depth \\ \hline 
Object            & 12 & 0 & 1 & 0 \\
Throwable         & 17 & 5 & 4 & 1 \\
Exception         & 0 & 1 & 4 & 2 \\
Component         & 254 & 80 & 1 & 1 \\
ComponentUI       & 11 & 0 & 1 & 1 \\
Container         & 106 & 18 & 1 & 2 \\
AbstractAction    & 12 & 3 & 3 & 1 \\
AccessibleContext & 24 & 24 & 1 & 1 \\
JComponent        & 178 & 69 & 1 & 3 \\
Expression        & 17 & 1 & 1 & 1 \\
RuntimeException  & 0 & 1 & 4 & 3 \\
Component.AccessibleAWTComponent & 39 & 2 & 1 & 2 \\
Buffer            & 21 & 5 & 1 & 1 \\
EventObject       & 2 & 1 & 1 & 1 \\
ORB               & 60 & 5 & 1 & 1 \\ \hline
\end{tabular}
\caption{\label{tbl:jdk-inheritance} 
	The fifteen classes with the highest inheritance PG values: JDK
}
\end{center}
\end{footnotesize}         
\end{table*}

We note that the JDK class {\tt java.awt.Component} with 254 methods
(Table~\ref{tbl:summary}) was ranked 25th in terms of reverse aggregation,
26th by aggregation and 5th by inheritance. This places the class
in the top 1\% for all metrics. We would therefore consider this a
key class. Additionally, it may be considered a {\em Large Class} \cite{FOWL99}
and hence a candidate for refactoring via {\em Extract Class} or
{\em Extract Subclass} because of this ``bad smell''.

Finally, it is also noticeable that six classes with the highest
reverse aggregation PG values have three or more constructors. This
would make those classes eligible for refactoring by {\em replacing
multiple constructors with creation methods} as suggested by Kerievsky
and empirically investigated by Najjar et al. \cite{KERI02,NAJJ03}.

Considering conjecture C1, the classes most often used in aggregation
relationships do not necessarily contain the highest numbers of methods
or attributes. Neither is the nature of those classes suitable for
use by subclasses. For example, {\tt String} and {\tt Class} are final
and {\tt Hashtable} and {\tt Vector} are themselves subclasses of
{\tt Dictionary} and {\tt AbstractList}, respectively.

\subsubsection{The Tomcat system}  
 
Table~\ref{tbl:tomcat-revagg} shows, for the Tomcat system, the numbers of
methods, fields and constructors and the depth in the inheritance
hierarchy, for the top fifteen classes when ranked in
descending order by their reverse aggregation PG values.

\begin{table*}[htb]
\begin{footnotesize}
\begin{center}
\begin{tabular} {|l|l|r|r|r|r|} \hline
Classname  &  Methods  &  Attributes  & Constructors & Depth \\ \hline
Options & 15 & 0 & 0 & 0 \\
JspCompilationContext & 40 & 24 & 1 & 1 \\
JspServletWrapper & 5 & 8 & 1 & 1 \\
Logger.Helper & 10 & 4 & 3 & 1 \\
Logger & 36 & 18 & 1 & 1 \\
JspReader & 24 & 9 & 1 & 1 \\
ErrorDispatcher & 19 & 2 & 1 & 1 \\
StringManager & 7 & 2 & 1 & 1 \\
ServletWriter & 13 & 6 & 1 & 1 \\
Compiler & 10 & 9 & 2 & 1 \\
JspRuntimeContext & 15 & 10 & 1 & 1 \\
ServletEngine & 2 & 2 & 1 & 1 \\
Mark & 10 & 10 & 3 & 1 \\
Node.Root & 3 & 1 & 1 & 2 \\
ErrorHandler & 3 & 0 & 0 & 0 \\ \hline
\end{tabular}
\caption{\label{tbl:tomcat-revagg} 
The fifteen classes with the highest reverse aggregation PG values: Tomcat
}
\end{center}
\end{footnotesize}         
\end{table*}

From Table~\ref{tbl:tomcat-revagg}, it is interesting that the
mean inheritance depth of the fifteen classes is considerably smaller
than in the JDK. Only one class, {\tt Node.Root} extends any class other
than {\tt java.lang.Object}, compared with six in the JDK.
It is also interesting to note that no class in the top fifteen when ranked by
inheritance is present in either of the lists for top classes when
ranked by normal or reverse aggregation PG values.

\begin{table}[htb]
\begin{footnotesize}
\begin{center}
\begin{tabular} {|l|r|r|}
\hline
Class & Reverse PG & PG \\ \hline
JspCompilationContext & 2 & 3 \\
JspServletWrapper & 3 & 9 \\
JspReader & 6 & 6 \\
Compiler & 10 & 5 \\
ServletEngine & 12 & 10 \\
Mark & 13 & 12 \\ \hline
\end{tabular}
\caption{\label{tbl:tomcat-overlap} 
	Seven class appear in the top fifteen classes for both
	normal and reverse aggregation PG values: Tomcat
}
\end{center}
\end{footnotesize}         
\end{table}

We would expect the depth values for both Tomcat and Ant to be smaller than
those for the JDK, and this is borne out in the values from Table~\ref{tbl:tomcat-revagg}.
A contributing factor to this is the role that the JDK class {\tt java.lang.Object}
plays - every class in the JDK extends class {\tt java.lang.Object} by default.

The top fifteen classes were then extracted for normal aggregation PG 
values. Table~\ref{tbl:tomcat-overlap} shows that six of these top fifteen could also be
found in the fifteen reverse aggregation values from Table~\ref{tbl:tomcat-revagg}.

We note that the class {\tt JspC} with 57 attributes (see Table~\ref{tbl:summary})
did not figure in either the highest normal or reverse aggregation PG 
values. Inspection of the raw data revealed it to be ranked 172 (reverse
aggregation) and 73 (normal aggregation) from 321 classes analysed. 
Visual inspection revealed that most attributes of the class were {\tt String}s
and {\tt int}s. Also, the class with 168 methods (see Table~\ref{tbl:summary}),
{\tt catalina.core.StandardContext}, has 49 attributes and is ranked 37th by
reverse aggregation PG, 14th by aggregation PG and 41st by inheritance.
Although it does not fit in the top fifteen by reverse aggregation, it may be
considered a candidate class for refactoring. The most obvious refactorings
relate to a bad smell - namely {\em Primitive Obsession} \cite{FOWL99}.

With regards to conjecture C1, it is not true to say that classes with substantial amounts 
of aggregation (either normal or reverse), or indeed, with substantial amounts of 
methods and/or attributes tend to have the highest number of descendents (i.e., 
tend to be found near the root of an inheritance hierarchy).

With respect to refactoring of constructors, we note that only two classes of those
listed in Table~\ref{tbl:tomcat-revagg} have three or more constructors. This
would imply less scope for refactoring by {\em replacing multiple constructors
with creation methods} in this system. This constrasts with the JDK where there
were six such classes.

\subsubsection{The Ant system}

\begin{table*}[htb]
\begin{footnotesize}
\begin{center}
\begin{tabular} {|l|l|r|r|r|r|} \hline
Classname  &  Methods  &  Attributes  &  Constructors & Depth \\ \hline
Path & 25 & 2 & 2 & 3 \\
Location & 1 & 4 & 3 & 1 \\
FileUtils & 26 & 4 & 1 & 1 \\
Project & 89 & 33 & 1 & 1 \\
FilterSet & 16 & 5 & 2 & 3 \\
FilterSetCollection & 3 & 1 & 2 & 1 \\
InputHandler & 1 & 0 & 0 & 0 \\
Commandline & 23 & 3 & 2 & 1 \\
Task & 23 & 8 & 1 & 2 \\
Target & 21 & 7 & 1 & 1 \\
RuntimeConfigurable & 11 & 6 & 1 & 1 \\
Compatibility & 1 & 1 & 1 & 1 \\
UnknownElement & 12 & 3 & 1 & 3 \\
SelectorUtils & 9 & 1 & 1 & 1 \\
CommandlineJava & 26 & 7 & 1 & 1 \\ \hline
\end{tabular}
\caption{\label{tbl:ant-revagg}
The fifteen classes with the highest reverse aggregation PG values: Ant
}
\end{center}
\end{footnotesize}         
\end{table*} 

Table~\ref{tbl:ant-revagg} shows values the the fifteen highest reverse aggregation 
values.
None of these classes were in the top fifteen classes when sorted by
aggregation PG values. Only one class, {\tt Task}
was present in the top fifteen when sorted by inheritance PG values.
This class was ranked 30th in the aggregation list - just outside
the top 1\%.
One possible explanation for the lack of overlap is that the inheritance
hierarchy in Ant centers around the {\tt Task} and {\tt ProjectComponent}
classes. This means that the other classes in Ant depend more upon
inheritance and less upon aggregation and delegation.

One hypothesis to explain this behaviour is that aggregation is used as a
surrogate for inheritance. The results from the JDK and Tomcat systems,
where there was very little overlap between inheritance and aggregation
PG values, and in particular from the Ant system, suggests that
this may be the case. The hypothesis that delegation is used as
a surrogate for multiple inheritance fits well with the observation
that the distribution of interfaces follows a power-law \cite{WHEE03c}.
Whilst many classes implement a few interfaces, a few classes implement a large
number of interfaces. Those that do, tend to delegate the responsibility for the
methods of these interfaces to members of the same interface.

In the Ant system, we would consider the top ten classes found by
ordering reverse aggregation PG value to be key classes.
Of these, {\tt Project} with 89 methods and 33
attributes stands out as {\em Large Class} and a possible candidate for
refactoring. Hence, conjecture C1 would seem to be supported for the
Ant system.

With regards to refactoring, only one class of those in Table~\ref{tbl:ant-revagg}
has three constructors. The rest are ineligible for refactoring by
{\em replacing multiple constructors with creation methods}. Together
with the result for Tomcat, this implies a significant difference
between libraries and application systems; namely key classes in
library systems tend to have a greater number of constructors.

\section{Meta-analysis of refactorings}

The focus of the research in this paper has been to identify candidate classes
for refactoring - i.e. key classes. We have chosen specific 
properties to identify such classes. We now need to justify why, for example,
we chose aggregation and inheritance relationships, together with the number
of methods and attributes as those features (as opposed to any other class 
features).  

To illustrate why, and as part of the research herein, a dependency diagram
showing the relationships between the seventy-two refactorings outlined in
Fowler et al. \cite{FOWL99} was developed. From this meta-analysis emerged
various {\em core} refactorings. By core refactorings, we mean those
refactorings upon which a large number of other refactorings depend 
and are required in each of those refactorings.  
The three most {\em important} of these, in
terms of the number of refactorings dependent on them, as 
established by the meta-analysis, are:
\begin{enumerate} 
\item {\bf Extract Method} - which should be performed when the code body of a
method is getting too long. 

\item {\bf Move Field} - which should be performed when that field is being
used by another class more than by the class in which it is defined.

\item {\bf Move Method} - which should be performed whenever a method is using
features of another class more than those in which it is defined. 
\end{enumerate} 

For both of the latter two core refactorings, and to a lesser extent
{\em extract method}, aggregation plays a central role. If we wish to move a
field, then the type of that field must be considered; it may be that of
another class. If we want to move a 
method from one class to another, then all types of coupling in that method
(including inheritance and aggregation) must be considered; removing 
the coupling 
from the source class may simplify it, but their loss needs to 
be replaced with appropriate code. 
Analysis of the mechanics of the two refactorings reveals inheritance 
to play a large role, in keeping with many other refactorings. 
For example, the Move Field refactoring requires as part of 
its mechanics that if the field 
is not declared as private, then all subclasses need to be 
checked for references to that
field. The Move Method refactoring 
requires all 
subclasses and superclasses
to be checked for references to that method. 
 
We also believe that
classes with large numbers of methods and attributes are likely to require 
application of these two refactorings at some point. 
Moreover, we refer to two common bad smells in classes 
identified in this paper from the key classes 
chosen - that
of {\em Large Class} characterised by too many methods and {\em Primitive
Obsession} characterized by overuse of primitive attributes and system classes.
We therefore justify our choice of criteria for selection of 
key classes on this basis of these arguments.

\section{Conclusions and Future Research} 

In this paper, we have described a technique by which key classes can
be identified from three Java systems. A metric, potential gain, was adapted
from use in Web search and navigation problems to evaluate the dependence
which the system places on various classes. By computing values for potential
gain on graphs for two forms of coupling, namely, inheritance and aggregation,
we were able to identify candidate key classes. By combining these results
with two other class features (number of class methods and class attributes)
we were able to further isolate classes in possible need of refactoring.

Four principal results emerged from the research. Firstly, that metrics
from other research domains can be adapted to aid developers and researchers
in the refactoring process.

Secondly, there are substantial differences between each of the three systems
investigated. The JDK system has key classes with more constructors. This
is reflected in the mean number of constructors for classes within these systems - 1.242 for
JDK against 1.073 for Ant and 1.069 for Tomcat.

Thirdly, there is also evidence to suggest that only the Ant system exhibits the
expected properties of key classes according to conjecture C1.
We therefore reject the hypothesis that key classes are consistently
found at the base of inheritance hierarchies.

Finally, the analysis supports previous work on power-laws, which suggested
that interfaces were commonly used as a surrogate for multiple implementation
inheritance.

Future work will focus on two areas. Firstly, expanding the scope of what a key
class is. The PG metric will be used to extract details relating to method
parameters, method return types and interfaces. A second area of future research
will investigate the potential for the key classes identified in this paper to be 
refactored. The research thus represents a first step in establishing the features 
of classes most eligible for refactoring. 

\bibliography{../bib/papers.bib}
\bibliographystyle{abbrv} 

\appendix

\section{Potential Gain}

Formally, if the fraction of all possible paths in a directed graph, $G=(N,E)$,
to a depth, $d > 0$, which start from a node, $n \in N$ is given by
\begin{displaymath}
	R_d(n) = \sum_{y \in Out(n)} \frac{R_{d-1}(y)}{\sum_{j \in N} R_{d-1}(j)}
\end{displaymath}
where $R_0=1$ and $N$ is the set of nodes, $E$ is the set of edges and
$Out(i)=\{j | (i,j) \in E\}$, then the Potential Gain of $n$ is given by
\begin{displaymath}
	Pg(n) = \sum_{k=1}^{d_{max}} R_k(x) f(x)
\end{displaymath}
The constant, $R_0=1$, whilst not strictly accurate, is used to to ensure that
$\log Pg(n) > 0$ holds for all $n$. 
Two reasonable functions for $f(d)$ are the reciprocal function:
\begin{displaymath}
	f(d) = reciprocal(d) = 1/d
\end{displaymath}
and the exponential decay function:
\begin{displaymath}\label{eq:discount}
	f(d) = decay(d) = \gamma^d
\end{displaymath}
where $0 < \gamma < 1$ is a constant. 

The original justification for the use of these measures in the context of
Web search was based on the assumption that the utility of browsing a page
diminishes with the distance of the page from the starting URL. This assumption
is consistent with experiments carried out on real Web data \cite{HUBE98,LEVE00b},
and with studies showing that the probability of a user following a path of
length $n$ decreases as $n$ increases \cite{SILV99}.

The justification for these measures in the context of coupling is based
on the fact that the influence that two classes have on each other diminishes
as the distance (in terms of coupling) increases. This is a features of many
patterns - such as {\em Mediator} and {\em Facade} \cite{GAMM95} which reduce
communication and dependencies by introducing ``middle-men''.

The PG metric was collected automatically for each type of coupling for each
class in the three systems, and the reciprocal function was used to compute
the potential gain values.

\end{document}